\documentclass[review]{elsarticle}
\usepackage{graphics}
\usepackage{graphicx}
\usepackage{float}
\usepackage{bm}
\usepackage{lineno,hyperref}
\modulolinenumbers[1]
\usepackage{siunitx}         
\graphicspath{ {./Figures/} } 
\usepackage{soul}
\usepackage{amsmath}
\usepackage{geometry} 
\usepackage{color} 
\usepackage{outlines}
\journal{arXiv}

\begin{document}

\begin{frontmatter}
	
	\title{An \emph{In Situ} Study of the Role of Pressure on Fe Recrystallization and Grain Growth during Thermomechanical Processing}
	
	\author[1]{Darren C. Pagan\corref{mycorrespondingauthor}}
 	\ead{dcp5303@psu.edu}
    \author[1]{Lukas A. Kissell}
    \author[2]{Matthew L. Whitaker}

	\cortext[mycorrespondingauthor]{Corresponding author}

	\address[1]{Materials Science and Engineering, The Pennsylvania State University, University Park, PA 16802}
    \address[2]{Mineral Physics Institute, Department of Geosciences, Stony Brook University, Stony Brook, NY 11794}

	\begin{abstract}

Elevated pressures are encountered in many metal forming processes that can alter microstructural evolution rates. Here we measure rate changes with pressure in recrystallization and grain growth in Fe through adaptation of synchrotron-compatible multi-anvil presses, originally designed for study of the mantle. Recrystallization and grain growth are monitored \emph{in situ} using high-energy X-ray diffraction.  Principal component analysis applied to the diffraction images is used to quantify evolution rates, with increasing pressure significantly slowing the process.  

	\end{abstract}

\end{frontmatter}





Elevated pressures are often encountered in many industrially relevant forming processes including forging, rolling, and extrusion that can reach several times the yield strength of an alloy \cite{hosford2011metal} along with direct pressure application during hot-isostatic pressing \cite{atkinson2000fundamental}. These elevated pressures will naturally influence the thermodynamics and kinetics of microstructural evolution in tandem with changes in temperature. Despite extensive literature documenting significant changes in mechanical properties like yield strength under applied pressure \cite{spitzig1976effect,spitzig1984effect}, this pressure dependence of mechanical response is rarely accounted for in modeling efforts. Even less well investigated are pressure's effects on microstructural evolution. As we move towards more sustainable alloy production processes that utilize less energy and emit less CO$_2$, understanding and controlling pressure's role on microstructural evolution can be a critical component of the effort.

While the role of pressure on mechanical response has been examined for nearly a hundred years \cite{lewandowski1998effects} starting with Bridgman \cite{bridgman1912v,bridgman1916effect,bridgman1952studies}, studies of the role of positive pressure with respect to microstructural evolution during processing are more limited. From a thermodynamics perspective, increasing pressure will change the free energy to drive microstructural evolution in metallic alloys \cite{chen2022thermodynamic}, while in the context of kinetics, studies have shown that substitutional diffusion is generally slowed under pressure \cite{weyland1971effect,decker1977diffusion}. Regarding recrystallization, a relatively early study on pure Cu indicates a slowing of recrystallization with application of pressure \cite{tanner1962effect} and similarly on $\alpha$ brass \cite{kuhlein1988influence}, but these studies relied on \emph{ex situ} characterization to estimate recrystallization rates.

Although \emph{in situ} diffraction measurements during heat treatment have become relatively common over the past two decades \cite{rocha2005fast,elmer2007direct,esin2014situ,ebner2019austenite,warchomicka2019situ,brown2021evolution,rodrigues2021effect}, \emph{in situ} measurements during application of temperature \emph{and pressure} relevant to common thermomechanical processing are generally not performed. A major challenge for studying microstructural evolution under pressure is that the devices are difficult to safely construct, particularly with inclusion of \emph{in situ} monitoring. Fortunately, the mineral physics community has developed synchrotron-X-ray-compatible loading devices for studying rock deformation in the lower crust and upper mantle which apply complex combinations of pressure, deviatoric stress, and high temperature. Here we demonstrate that these same facilities can be extended to study  \emph{in situ} microstructural evolution in conditions simulating hot metal forming processes. The capability is demonstrated through \emph{in situ} X-ray diffraction measurements of slowed recrystallization and grain growth in pure Fe with increasing pressure applied. Extraction of transformation rates from the diffraction data is facilitated by machine-learning dimensionality reduction.

\begin{figure}[h]
      \centering \includegraphics[width=0.9\textwidth]{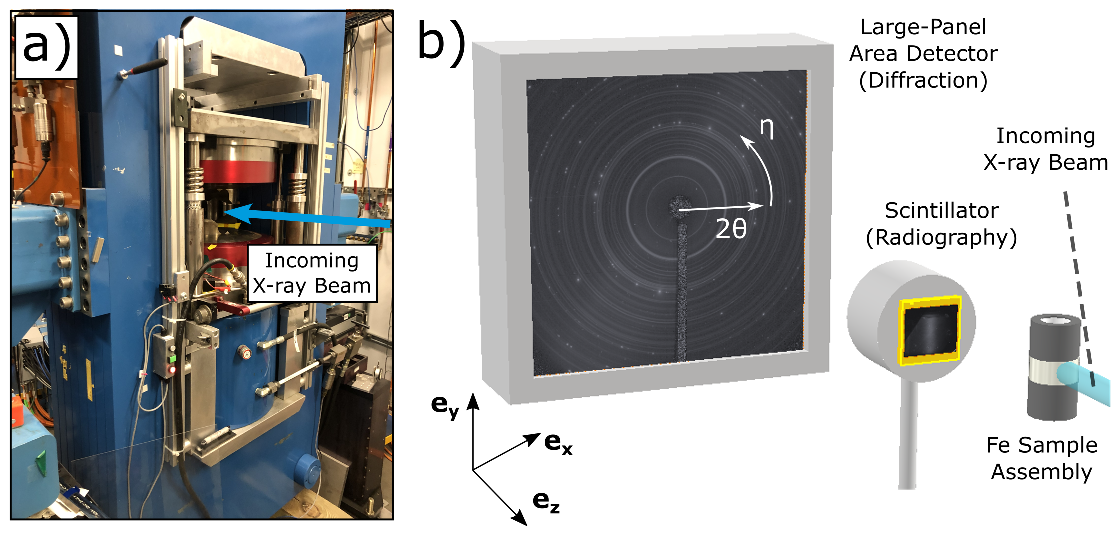}
      \caption{a) Image of the X-ray compatible hydraulic loading press used for the high-pressure recrystallization and grain growth studies. b) Schematic of the detectors used for \emph{in situ} X-ray measurements. A scintillator collecting the direct beam was used to measure macroscopic strain from X-ray radiographs, while a large-panel diffraction detector further from the specimen was used to monitor recrystallization and grain growth.}
      \label{fig:exp_setup}
\end{figure}

\begin{figure}[H]
      \centering \includegraphics[width=0.6\textwidth]{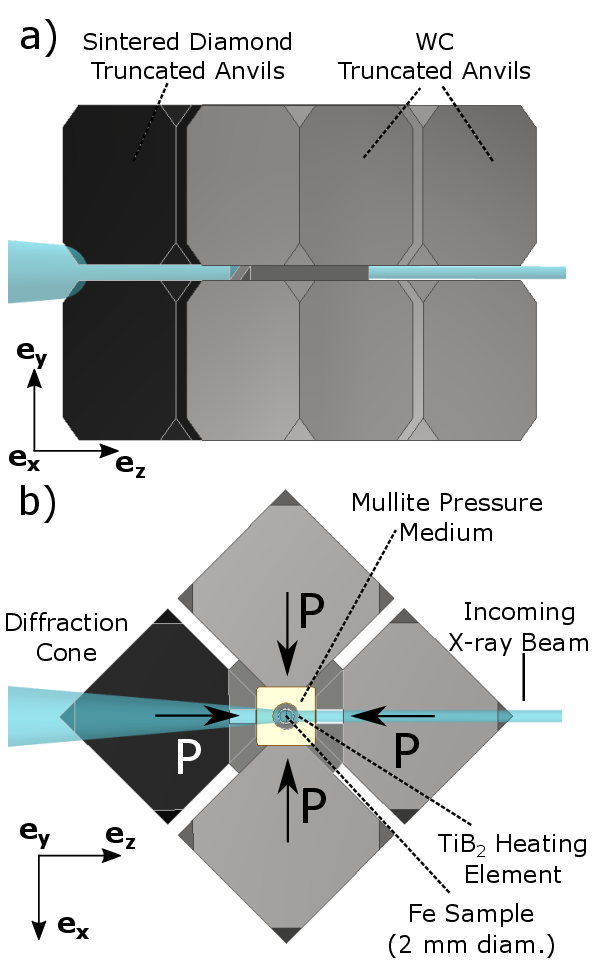}
      \caption{a) Side view of the Kawai-style anvil geometry used to compress the Fe specimens. b) Top-down cross-sectional view of the Kawai-style anvil geometry.}
      \label{fig:assembly}
\end{figure}

To study the effects of pressure on recrystallization and grain growth \emph{in situ}, high-pressure measurements were performed on the X-ray Powder Diffraction beamline (XPD, 28-ID-2) at the National Synchrotron Light Source II \cite{laasch2022outer}. An image of the Rockland Research hydraulic press in the beamline used for testing is shown in Fig. \ref{fig:exp_setup}a, while a schematic of the X-ray measurement geometry is given in Fig. \ref{fig:exp_setup}b. The specimen was illuminated with a 68.1 keV X-ray beam traveling in the -$\bm{e_z}$ direction. For imaging measurements, the beam was 2.2 mm $\times$ 2.2 mm (horizontal by vertical) to fully illuminate the specimen. The beam size was reduced to 300 $\mu$m $\times$ 100 $\mu$m for diffraction measurements. Specimens were deformed within the hydraulic press using 8 truncated cubic anvils which are independently controlled with three hydraulic pumps to produce varied stress states. A primary pump applies a hydrostatic pressure while two differential pumps can vary the stress state along the vertical direction $\bm{e_y}$. The 8 anvils (2 sintered diamond and 6 WC) are arranged into a Kawai-style geometry \cite{yamazaki2019high,irifune2002application} shown in Fig. \ref{fig:assembly}a. The sintered diamond anvils are X-ray transparent and arranged downstream to allow diffraction from the specimen to be transmitted.   

The cylindrical specimens tested were extracted from a 99.99\% Fe rod (1 in diameter $\times$ 6 in long) attained from MilliporeSigma. The specimen dimensions were 2 mm diameter $\times$ 2 mm tall (along $\bm{e_y}$) to be compatible with the sample testing environment. A cross section of the anvil loading system and the full sample assembly is given in Fig. \ref{fig:assembly}b. The sample was heated using direct resistance heating through the WC anvils and a BN-doped TiB$_2$ sleeve placed around the specimen. The specimen and sleeve were then placed into a mullite (2Al$_2$O$_3$ SiO$_2$) solid pressure medium that transmits load from the anvils to the specimen.

The testing procedure consisted of first building a sufficient dislocation density within the specimens by deforming the specimens to 12\% macroscopic strain in uniaxial compression. After the specimens had been deformed, they were then pressurized to target pressures $P$ of 1 GPa and 2 GPa. Pressures were reached by applying prescribed tonnages on the hydraulic press (222 and 374 kN respectively) determined from precalibration of the system. Temperature in the specimens was then ramped to 675 $^\circ$C by increasing power through direct resistive heating (128 W target) to initiate recrystallization within the specimens. The choice of temperature was informed by previous measurements to determine a temperature level that would drive grain growth in a time frame commensurate with detector speed. Active pressure control was then enabled during heating to maintain a fixed pressure on the specimens during thermal expansion. Power was maintained to $\pm$ 3 W during heating, corresponding to $\pm$ 16 $^\circ$C.  The true specimen pressures and temperatures were calculated after testing from lattice strains determined from peak shifts in the diffraction data, before and after application of pressure and temperature. Translation from strain to pressure and temperature was performed using linear elasticity (bulk modulus of 164.7 GPa \cite{hughes1953second}) and linear thermal expansion (coefficient of thermal expansion of 11.5$\times$10$^{-6}$ K$^{-1}$ \cite{nix1941thermal}) calculations respectively. The true specimen pressures and temperatures determined from the diffraction data are given in Table \ref{tab:test_mat}. For brevity, Tests 1 and 2 will be referred to as the 1 GPa and 2 GPa tests respectively.

\begin{table}
\centering
\begin{tabular}{c|c|c|c|c}
& Target $P$ (MPa) & Actual $P$ (MPa) &  Target $T$ ($^\circ$C) & Actual $T$ ($^\circ$C) \\ \hline
Test 1 & 1000 & 963 & 675 & 675 \\ \hline
Test 2 & 2000 & 2097 & 675 & 685 \\ \hline

\end{tabular}
    	 \caption{Test matrix showing the target and actual pressures $P$ and temperatures $T$ for the two recrystallization and grain growth tests.}
    	 \label{tab:test_mat}
\end{table}

 The current macroscopic strain in the specimens during prestraining was measured using X-ray absorption radiography to monitor the top and bottom of the specimens in real time. Radiography was performed using a YAG scintillator placed 470 mm from the specimen. The scintillator was coupled to a PointGrey Grasshopper 3 camera (2448 $\times$ 2048 pixels with 3.45 $\mu$m pixel size) and a variable optical lens. Magnification was chosen such that the effective pixel size for the imaging was 1 $\mu$m, providing effective strain resolution of approximately $5 \times 10^{-4}$. Once the target temperature was reached, diffraction data was collected continuously at a rate of 1 Hz on a PerkinElmer XRD 1621 detector sitting 1318 mm behind the specimen. The detector has 2048 $\times$ 2048 pixels with a size of 200 $\mu$m. In total, 1920 images were collected during the 1 GPa test and 6560 images during the 2 GPa test. After heating, the specimen was quenched and depressurized.  We note that uncontrolled specimen deformation that occurs during depressurization prevented preservation of the specimens' microstructure for post-testing microscopy.



\begin{figure}[H]
      \centering \includegraphics[width=1.0\textwidth]{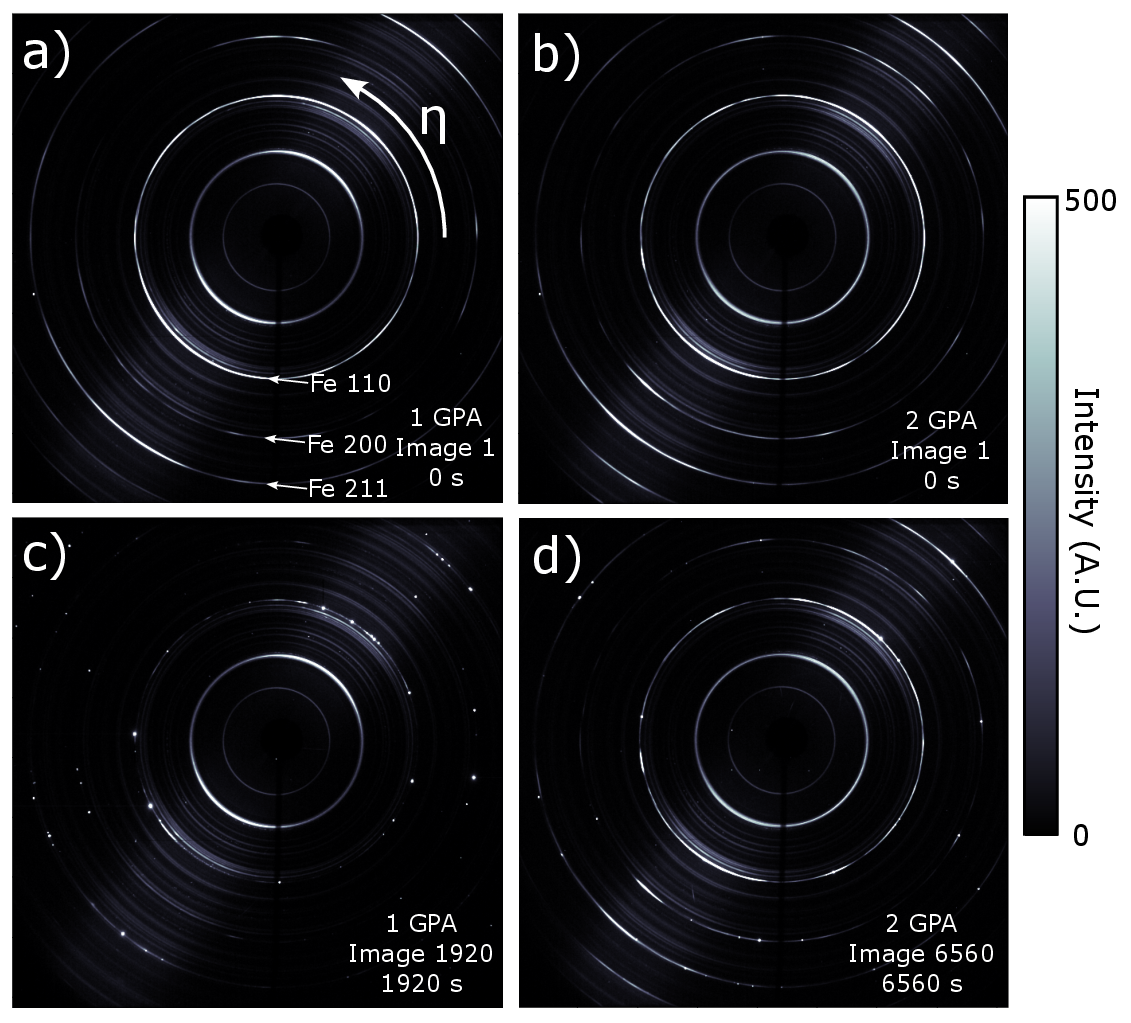}
      \caption{Diffraction images at the beginning of the experiments at 675 $^\circ$C with pressures of (a) 1 GPa and (b) 2 GPa applied. Diffraction images after the recrystallization tests have completed with (c) 1 GPa and (d) 2 GPa applied. The bright, isolated diffraction spots that form during the tests are emitted from large grains with low defect content that form during recrystallization and grain growth. The azimuthal angle $\eta$ on the detector has been labeled.}
      \label{fig:images}
\end{figure}

Figure \ref{fig:images} shows the evolution of the diffraction patterns at the beginning and end of specimen heating for the 1 and 2 GPa tests. Initial diffraction patterns for the 1 and 2 GPa tests are provided in Figs. \ref{fig:images}a and \ref{fig:images}b respectively, and final diffraction patterns are provided in Figs. \ref{fig:images}c and \ref{fig:images}d. The azimuthal angle $\eta$ has also been labeled. Diffraction rings from the (110), (200), and (211) sets of lattice planes are labeled in Fig. \ref{fig:images}a, but diffraction rings are complete only for the (110) and (200) lattice planes. Other diffraction rings present are from the BN-doped TiB$_2$ sample sheathing and mullite pressure medium. The intensity scaling in the figure is chosen to make the diffracted intensity from the Fe clearly visible. In Figs. \ref{fig:images}a and \ref{fig:images}b, at the beginning of the temperature soak, the Fe diffraction rings have non-uniform intensity around the ring, but the intensity is generally continuous which indicates a fine-grain microstructure or grains with large amounts of misorientation in addition to preferred orientation within the specimens. After the temperature soak (Figs. \ref{fig:images}c and \ref{fig:images}d), strong isolated diffraction peaks are observed at the positions of the Fe diffraction rings. These diffraction peaks correspond to newly formed, large grains with low defect content. This transition from continuous diffraction rings to isolated peaks is representative of the recrystallization process.

Quantitative analysis of this recrystallization and grain growth process (i.e., transition from continuous rings to discrete spots in the data) is a challenge in comparison to more common quantitative analysis of phase transformations using Rietveld refinement \cite{young1993rietveld}. For this reason, we utilize principal component analysis (PCA) to extract quantitative measures of the rates of the recrystallization and grain growth process at pressure. PCA is a form of  unsupervised machine learning \cite{van2009dimensionality} in which high-dimensional data (here the tens of thousands of pixels of diffracted intensity around Fe rings of interest) are projected along orthogonal `principal directions' of maximum variance of the dataset to obtain `scores' ($z_i$). Previous efforts have demonstrated that lower-dimensional descriptors of diffraction data directly reflect changes in the microstructure being probed \cite{pagan2019unsupervised,pagan2020informing,shadle2022using}. In this case, the lower-dimensional descriptors reflect the formation and growth of new grains and the corresponding texture changes  (as seen in the transition from continuous diffraction rings to distinct diffraction spots). The principal directions are the eigenvectors of the covariance matrix of each dataset. In most datasets, a relatively small number of projections or scores can capture a majority of the variance contained, thereby reducing the dimensionality. The eigenvalues of the covariance matrix reflect the total variance captured by each principal component direction.

To minimize contributions from the other diffracting materials in the X-ray beam path, diffracted intensity from around the {110} and {200} diffraction rings (the two complete diffraction rings) was utilized in the PCA. A more detailed explanation of this peak extraction procedure prior to dimensionality reduction can be found in \cite{pagan2019unsupervised}. Here, the sets of diffraction images from the 1 GPa and 2 GPa pressure tests are treated as separate datasets with different corresponding principal directions. For both tests, a majority of the data variance is described by projecting the data along the first two principal components. As such, we focus on the first two principal component scores, $z_1$ and $z_2$. 

\begin{figure}[H]
      \centering \includegraphics[width=0.6\textwidth]{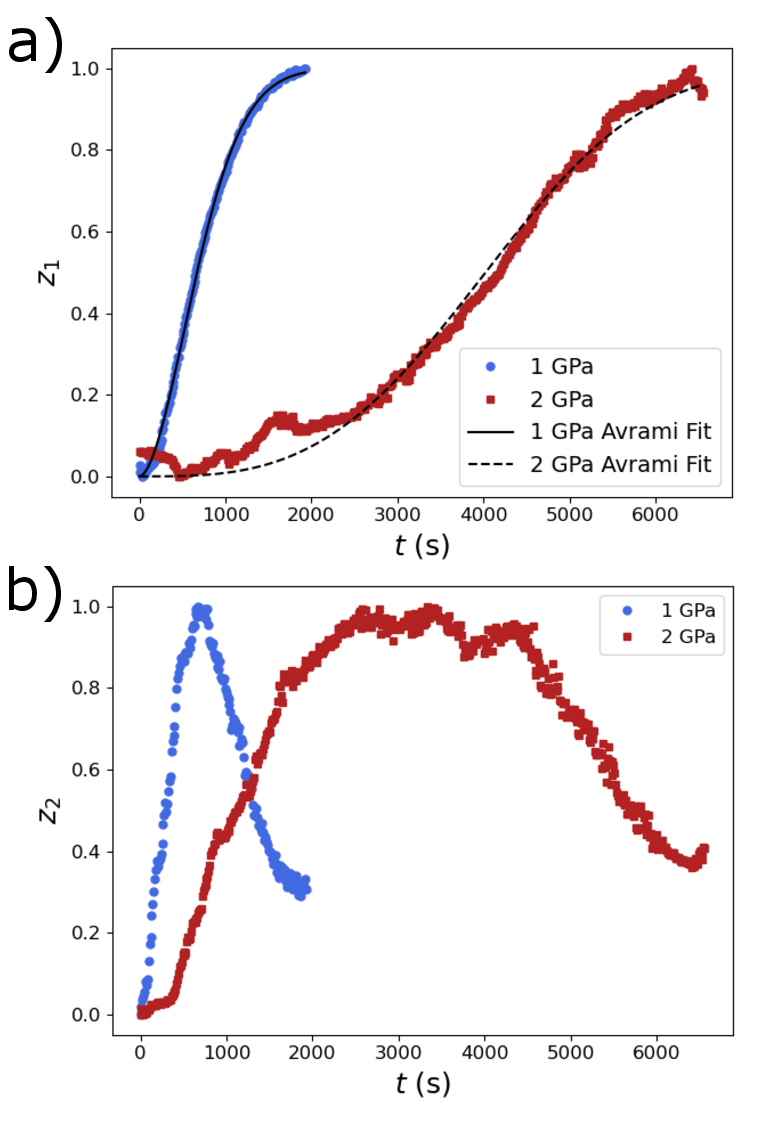}
      \caption{a) Time $t$ evolution of the first PCA score $z_1$ with time for the 1 (blue) and 2 (red) GPa tests. JMAK functions have been fit to $z_1$ for both tests. b) Time $t$ evolution of the second PCA score $z_2$ with time for the 1 (blue) and 2 (red) GPa tests.}
      \label{fig:pca}
\end{figure}

\begin{figure}[H]
      \centering \includegraphics[width=0.75\textwidth]{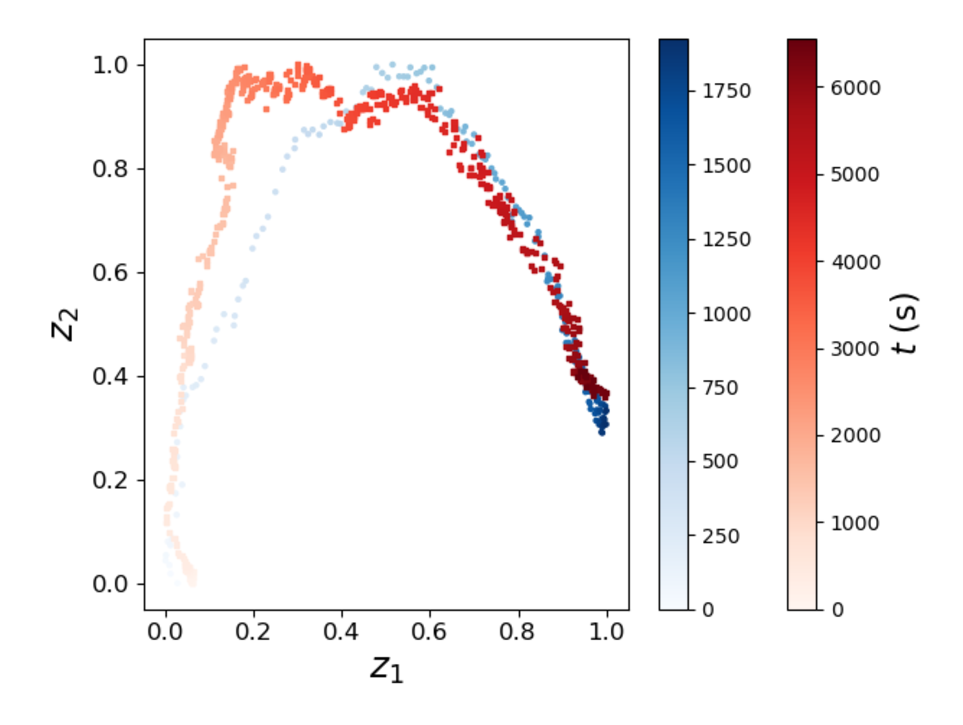}
      \caption{Diffraction data from the recrystallization experiments at 1 (blue) and 2 (red) GPa plotted in a lower-dimensional space as determined from PCA. Lower-dimensional data is colored according to time $t$.}
      \label{fig:pca2}
\end{figure}

The first two principal components capture 71.5\% and 9.6\% respectively of the 1 GPa data variance (81.1\% total) and 52.9\% and 18.4\% respectively of the 2 GPa data variance (71.3\%). Figure \ref{fig:pca}a shows the evolution of the first PCA score $z_1$ with time $t$ from the two recrystallization tests. The scores from the 1 GPa test are plotted in blue, while scores from the 2 GPa test are plotted in red. In this and other portions of Fig. \ref{fig:pca}, each marker corresponds to a score extracted from a diffraction image. For clarity, every 10\textsuperscript{th} score is shown.  We note that scale and sign of the scores are arbitrary, so in this work, scores are rescaled between 0 and 1. The time evolution of $z_1$ in both tests shows a monotonic increase and then a plateau at the end of the test. Figure \ref{fig:pca}b shows the time evolution of the second score $z_2$. The second score $z_2$ in both tests exhibits an increase to a maximum and then a decrease to less than half of the maximum. Figure \ref{fig:pca2} displays both data sets in a lower-dimensional space by projecting along the first two principal components. In Fig. \ref{fig:pca2}, the marker color corresponds to time through the recrystallization tests. In the lower-dimensional space, the diffraction data and underlying microstructure take similar paths, starting and ending in similar positions. However, as can be seen in all portions of the figure, the recrystallization process takes significantly longer under increased pressure (nearly $4 \times$).


While determining which exact features are being captured by the PCA is challenging, the time evolution of the first PCA score $z_1$ appears to reflect the relative level of completion of the recrystallization and grain growth process. This interpretation is supported by the observation that $z_1$ evolution in both tests exhibits a characteristic sigmoidal shape. In addition, a majority of the variance in both data sets is captured solely by the first PCA score. With this in mind, the crystallization time $t_c$ of the 1 GPa test is approximately 1500 s, and is nearly 6000 s for the 2 GPa test. This slowing of the recrystallization and grain growth process is consistent with previous observations that pressure slows substitutional diffusion and, in turn, grain growth \cite{weyland1971effect,decker1977diffusion}. Interpretation of the time evolution of the second PCA score $z_2$ is less clear, but it is apparent from the difference in $z_{2}$ evolution between the two pressures that a larger fraction of recrystallization time is spent at the maximum for 2 GPa compared to 1 GPa.

\begin{figure}[h]
      \centering \includegraphics[width=0.9\textwidth]{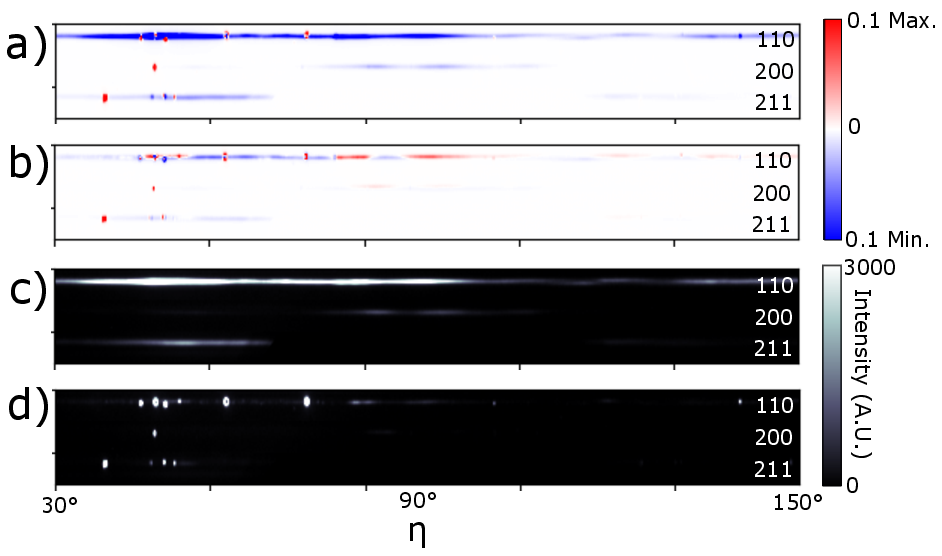}
      \caption{a) First and b) second principal components extracted from the 1 GPa test data. Example c) initial and d) final diffraction data which is projected on to the principal components. Only 120$^\circ$ of the full 360$^\circ$ about $\eta$ of the peak data is shown so features can be discerned. }
      \label{fig:comps}
\end{figure}

We can further explore the features that are influencing the scores by directly inspecting the principal components themselves. Figures \ref{fig:comps}a and \ref{fig:comps}b show a 120$^\circ$ azimuthal region of the first and second principal components respectively extracted from the 1 GPa test data. The shorter azimuthal range allows example features in the components to be more clearly discerned. Figures \ref{fig:comps}c and \ref{fig:comps}d shows example data from the start and end of the 1 GPa test for reference. The first principal component has contributions from both the initial and final diffraction data configurations. When projecting the data against this principal component, the continuous ring data at the start of the test will produce a lower score while the distinct peaks at the end of the test will produce a higher score. This essentially means the first score captures the degree of transformation transitioning from low to high. The second principal component is much more dominated by the distinct diffraction peaks that appear during recrystallization and grain growth. There even appears to be peaks that are not present in the final configuration which reflect grains that were either consumed or rotated out of the diffraction condition. Projecting the data against this principal component will produce a score that reflects the number of grains with low defect content that are currently diffracting to produce distinct diffraction peaks along the ring.

Furthermore, with the interpretation that the first PCA score reflects the degree of completion of the microstructure evolution process, a Johnson-Mehl-Avrami-Kolmogorov (JMAK) function \cite{kolmogorov1937statistical,avrami1939kinetics,avrami1941granulation,william1939reaction} was fit to the evolution of the first PCA scores $z_1(t)$ to explore the rates of the transformation process:
\begin{equation}
    z_1(t)=1-e^{-Kt^n}
\end{equation}
where $K$ and $n$ are fitting parameters. The fits of the JMAK function to the evolution of $z_1$ for both tests are shown in Fig. \ref{fig:pca}a. Interestingly, the fit exponents $n$ of the two tests change fairly dramatically from $n=1.8$ (approximately 2) for the 1 GPa test to $n=3.2$ (approximately 3) for the 2 GPa test. The exponent $n$ has been related to the characteristics of nucleation site distribution and the nucleation rate \cite{cahn1956transformation}. Therefore, this increase in exponent $n$ with pressure may be related to an increase in nucleation sites at higher pressure, despite the much slower crystallization time. This may be rationalized by the fact that pressure will increase the local strain energy to initiate the recrystallization, leading to more nucleation events. Interestingly, this may be connected to the evolution of the second PCA score $z_2$ (Fig. \ref{fig:pca}b) where the 2 GPa test exhibited an extended fraction of the crystallization time at peak value. If $z_2$ is connected to nucleation events, then more of the crystallization time at peak value would coincide with a larger exponent $n$ fit to the recrystallization process. However, further studies would be required to confirm this supposition.

While \emph{in situ} monitoring of high pressure processes is a challenge, here we have demonstrated that a promising approach for studying metal thermomechanical processing is to adapt synchrotron experimental capabilities originally designed for studying mineral physics. In this work, it was demonstrated that pressure significantly alters the rates of the recrystallization and grain growth process. The findings here indicate that pressures encountered during many standard thermomechanical processes will slow grain growth due to slowed substitutional diffusion in Fe (and likely steels) and increase the number of nucleation sites for recrystallization due to increased free energy. Together, these effects of pressure will lead to smaller grains and finer microstructures, partially contributing to the fine microstructures often found during severe plastic deformation processes at high pressure \cite{iwahashi1996principle,zhilyaev2003experimental}. More importantly, quantitative information regarding recrystallization and grain growth rates could be extracted from the complex transition from full diffraction rings to isolated spots using data dimensionality reduction (principal component analysis). As we attempt to increase the sustainability of metal alloy thermomechanical processing, understanding and quantifying the role of pressure on microstructural evolution rates is a means to reach target, or even new, microstructures without significantly increasing energy usage. The approach reported here presents a path forward for gathering the data necessary to develop more sustainable, heat-treatment-less, alloy processing routes across a wide range of alloy classes.





\section*{Acknowledgments}

DCP and LAK acknowledge the support of FIERF Grant \#258985. This research used resources 28-ID-2 of the National Synchrotron Light Source II, a U.S. Department of Energy (DOE) Office of Science User Facility operated for the DOE Office of Science by Brookhaven National Laboratory under Contract No. DE-SC0012704. Use of the MAXPD endstation was supported by COMPRES, the Consortium for Materials Properties Research in Earth Sciences, under NSF Cooperative Agreement No. EAR 16-61511 and by the Mineral Physics Institute, Department of Geosciences, Stony Brook University.

\section*{Conflict of Interest Statement}
On behalf of all authors, the corresponding author states that there is no conflict of interest.

\newpage
\clearpage

\bibliographystyle{elsarticle-num}
\bibliography{References.bib}

\end{document}